\documentclass[prb,twocolumn,reprint,showpacs,superscriptaddress,floatfix]{revtex4-1}
\usepackage{epsfig}
\usepackage[dvips]{color}

\begin{document}

\title{Crystal growth, resistivity and Hall effect of the delafossite metal PtCoO$_2$}

\author{Pallavi Kushwaha}
\email[corresponding author: ]{Pallavi.Kushwaha@cpfs.mpg.de}
\affiliation{Max Planck Institute for Chemical Physics of Solids, N$\ddot{o}$thnitzer Stra$\beta$e 40, 01187 Dresden, Germany}
\author{Philip J. W. Moll}
\affiliation{Laboratory for Solid State Physics, ETH Zurich, Schafmattstr. 16, CH-8093 Zurich, Switzerland}
\author{Nabhanila Nandi}
\affiliation{Max Planck Institute for Chemical Physics of Solids, N$\ddot{o}$thnitzer Stra$\beta$e 40, 01187 Dresden, Germany}
\author{Andrew P. Mackenzie}
\affiliation{Max Planck Institute for Chemical Physics of Solids, N$\ddot{o}$thnitzer Stra$\beta$e 40, 01187 Dresden, Germany}
\affiliation{Scottish Universities Physics Alliance, School of Physics and Astronomy, University of St. Andrews, St. Andrews KY16 9SS, United Kingdom}

\date{\today}

\begin{abstract}
We report single crystal growth of the delafossite oxide PtCoO$_2$, and basic transport measurements on single crystals etched to well-defined geometries using focused ion beam techniques.  The room temperature resistivity is 2.1 $\mu\Omega$~cm, and the Hall coefficient is consistent with the existence of one free electron per Pt.  Although the residual resistivity ratio is greater than fifty, a slight upturn of resistivity is seen below 15~K.  The angle dependence of the in-plane magnetoresistance is also reported. 

\end{abstract}
\maketitle

\section{Introduction}
The delafossite structural series of oxides has the general formula ABO$_2$, in which A is a noble metal (Pt, Pd, Ag or Cu) and B is a transition metal (e.g. Cr, Co, Fe, Al and Ni) [\onlinecite{Shannon, Rogers, Prewitt, Marquardt}].  The metal atoms are found in layers with triangular lattices, stacked along the perpendicular direction in various sequences, leading to structures similar to the layered rock salt structure adopted by the well-known ionic conductor LiCoO$_2$ and by NaCoO$_2$, which superconducts when intercalated by water.  Interlayer coupling is weak, so the delafossites are quasi-two dimensional, and the physical properties across the series vary considerably with A and B combinations.  Known materials include candidate magnetoelectric insulators and thermoelectrics [\onlinecite{Singh, Wang}], transparent semiconductors [\onlinecite{Yanagi}] and band insulators [\onlinecite{Shannon, Rogers, Prewitt, Marquardt}].  There are also a few intriguing metals. AgNiO$_2$ and PdCrO$_2$, for example, provide a rare opportunity to study the interplay between complex magnetic order and triangular lattice conductivity [\onlinecite{ Takatsu 2009, Takatsu 2010, Ok, Takatsu 2014, Coldea}].  However, magnetism is not necessary for a delafossite metal to be of interest. 

\setlength{\parskip}{5mm}
In recent years, non-magnetic PdCoO$_2$ has attracted considerable attention.  The cobalt has formal valence of $3+$ and a $3d^{6}$ configuration, meaning that the states at the Fermi level have dominantly Pd character [\onlinecite{Eyert, Seshadri, Kim}].  The Fermi surface consists of a single cylinder of hexagonal cross-section as determined by ARPES [\onlinecite{Noh}], and a volume corresponding to one electron per Pd as determined by dHvA measurements [\onlinecite{Hicks}].  The Fermi surface shape would suggest that it forms dominantly from Pd $4d$ states, but the conductivity is extremely high.  At 300~K the resistivity ($\rho$) is $2.6~\mu\Omega$~cm, higher per carrier than that of Cu, suggesting a prominent role for Pd $5s$ states [\onlinecite{Hicks}].  At low temperature both the shape and the value of $\rho$ are noteworthy.  It is essentially independent of temperature below 15~K, and is best fitted by $\rho$ = $\rho$$_0$ + $\beta$$e^{-T^{*}/T}$ (with $\beta$ a constant and $T^{*}$ = 165~K) rather than the standard power law expected for standard phonon scattering.  This can be explained semi-quantitatively by postulating the occurrence of strong phonon drag, meaning that electron-phonon scattering relaxes momentum only through Umklapp scattering, which is essentially gapped out by 15~K [\onlinecite{Hicks}].  At least as surprising as the form of $\rho$ is its value.  A residual resistivity of only $7.5~n\Omega$~cm has been reported in flux-grown crystals.  Even if phonon scattering has become ineffective because of phonon drag and a large Umklapp gap, it is hard to understand such a low value in as-grown crystals that have not been subjected to any post-growth purification.  It corresponds to a mean free path ($\textit{l}$~) of $20~\mu$m, or $\approx 10^{5}$ lattice spacings. 

\setlength{\parskip}{5mm}
The high value of $\textit{l}$ in PdCoO$_2$ has important consequences for a number of physical properties, for example the out of plane magnetoresistance which is huge and varies strongly with field angle [\onlinecite{Takatsu 2013}].  It also brings the material into an interesting regime in which the momentum relaxation rate as measured by the resistivity is slow.  There is at least the possibility that, under these circumstances, Boltzmann transport theory might break down in favor of a regime in which a hydrodynamic description [\onlinecite{Andreev, Gurzhi, Jong}] is more appropriate for the electron fluid.  An important question, therefore, is why PdCoO$_2$ is so special.  Signatures of phonon drag physics were difficult to observe in the resistivity of standard metals, but in PdCoO$_2$ a pronounced effect is seen.  In Ref. [\onlinecite{Hicks}] it is argued that a combination of dimensionality, lattice stiffness and Fermi surface topography mean that the characteristic temperature for Umklapp processes is high, enhancing the visibility of the $e^{-T^{*}/T}$ contribution to the resistivity.  However, it is important that this be checked by searching for other materials which, by the above arguments, should show similar behavior.  

\setlength{\parskip}{5mm}
The most obvious material as the first choice for such a comparison is the sister compound of PdCoO$_2$, PtCoO$_2$.  Although crystal growth [\onlinecite{Prewitt, Tanaka}] and a low room temperature resistivity [\onlinecite{Rogers}] were reported decades ago, no temperature-dependent transport studies were carried out on those crystals, and no further crystal growth has been reported since then.  In this paper we describe our growth and initial transport studies of single crystals of PtCoO$_2$.

\begin{figure}[t]
	\begin{center}
	\includegraphics[width=8.5 cm]{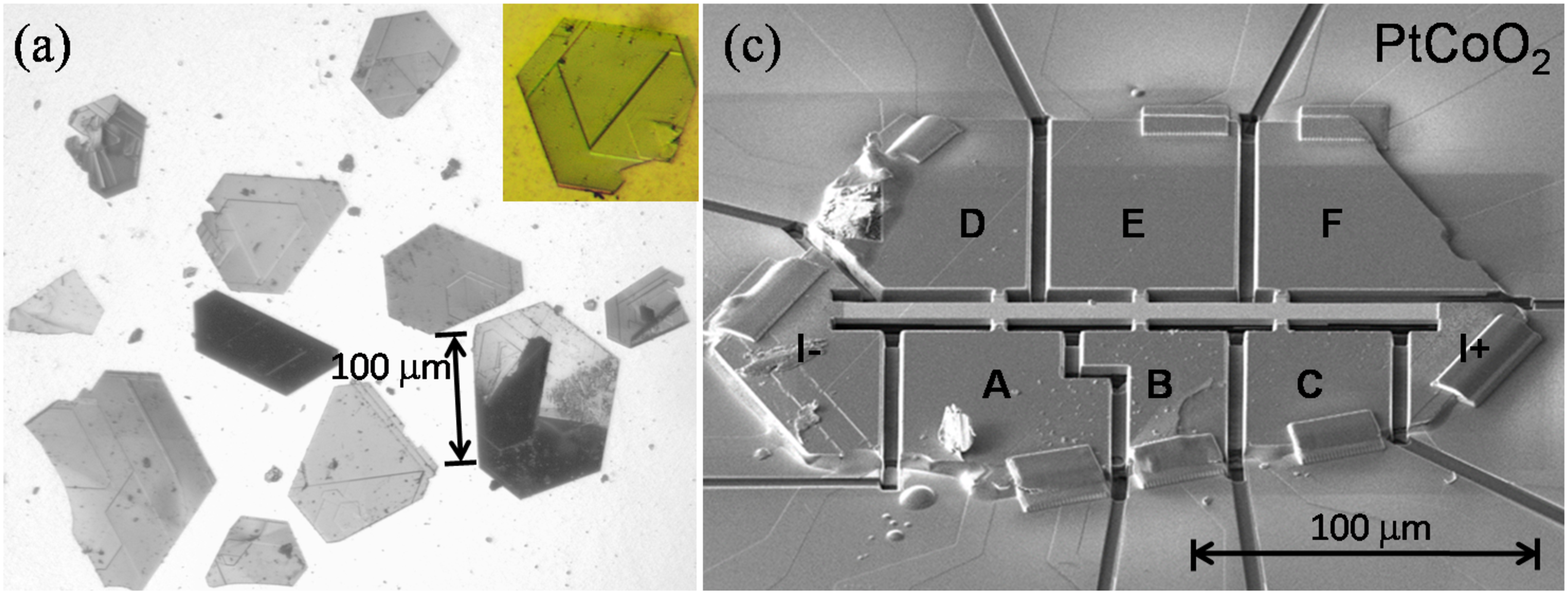}
	\includegraphics[width=8.5 cm]{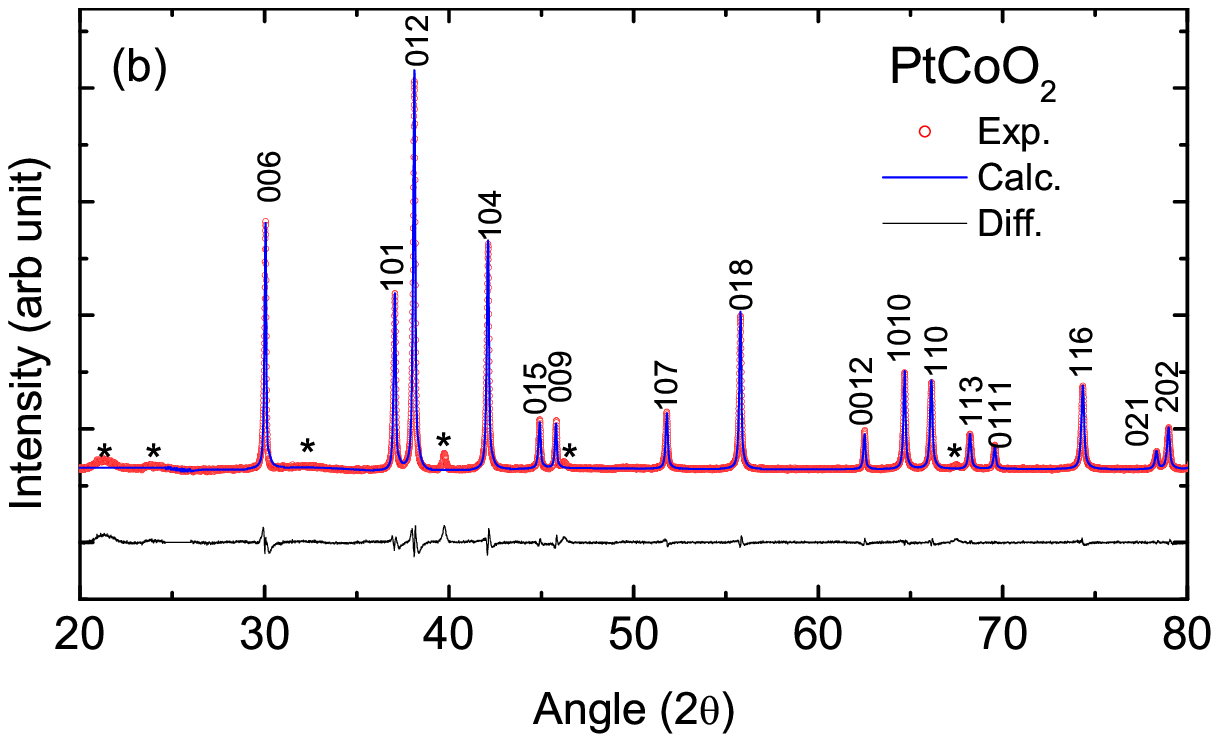}
	\end{center}
	\caption{(color online) (a) Optical microscope image of as-grown crystals of PtCoO$_2$. (b) Le-Bail fitting of powder x-ray diffraction pattern along with fitted curve, and the difference curve. All peaks are labeled with corresponding $\textit{hkl}$ values. Peaks marked with $\ast$ correspond to unavoidable unreacted PtCl$_2$ stuck to the crystal surface. (c) SEM image of a sample used for transport measurements in which a focused ion beam was used to define a measurement track of well-defined geometry.}
	\label{Figure1}
\end{figure}

\begin{figure}[b]
	\begin{center}
	\includegraphics[width=8 cm]{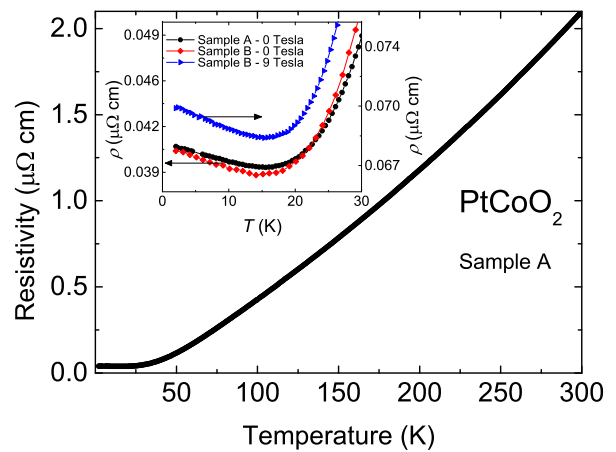}
	\end{center}
	\caption{(color online) The temperature dependent in-plane resistivity of PtCoO$_2$ in zero applied magnetic field. The inset shows details of the behavior below 30~K, highlighting the upturn in resistivity below 15~K.  Data from two crystals are shown in zero applied field, and from one in 9 T, demonstrating that the magnetoresistance is positive throughout the region in which the upturn is seen.}
	\label{Figure2}
\end{figure}

\section{Experimental details}
First, we describe our crystal growth.  Shannon \textit{et. al}  [\onlinecite{uspatent}] reported single crystal growth of Pt$_x$Co$_y$O$_2$ (where x and y were $0.85~\pm~0.15$) by using different techniques several decades ago.  Stoichiometric PtCoO$_2$ was grown only under high pressure (3000 atm). Other growth conditions resulted only in nonstoichiometric crystals. Tanaka \textit{et. al} [\onlinecite{Tanaka}] were able to grow stoichiometric PtCoO$_2$ by using a metathetical reaction under vacuum, but the crystal size was limited up to $30~\mu$m. Here we used a technique similar to that applied in Ref. [\onlinecite{Takatsu 2007}] to the growth of PdCoO$_2$. Using a modified temperature profile we first grew PdCoO$_2$ and then extended the technique to the growth of PtCoO$_2$.
\begin{figure}[t]
	\begin{center}
	\includegraphics[width=8 cm]{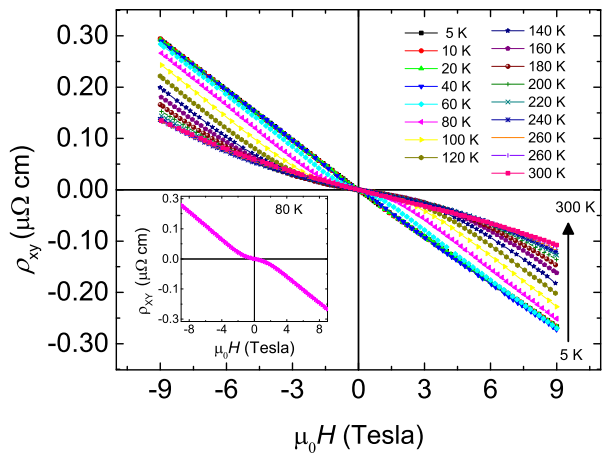}
	\end{center}
	\caption{(color online) Field dependence of Hall resistivity ($\rho_{xy}$) for PtCoO$_2$ for different labeled temperatures for the $2.6~\mu$m thick sample. The inset highlights the two regions seen as a function of applied magnetic field, using data taken at 80~K as an example.}
	\label{Figure3}
\end{figure}

Powder of reagent grade PtCl$_2$ ($99.99+\%$ purity, Alfa Aesar) and CoO ($99.995\%$ purity, Alfa Aesar) were ground together for approximately an hour under an inert atmosphere in accordance with the chemical reaction PtCl$_2$ + 2CoO$\rightarrow$ PtCoO$_2$ + CoCl$_2$. The mixed powder was then sealed in a quartz tube under a vacuum of $5\times10^{-6}$ Torr.  The sealed quartz tube was heated in a vertical furnace to 800~$^{\circ}$C for 5 hours and cooled down to 740~$^{\circ}$C at a rate of 7.5~$^{\circ}$C/hour and kept at this temperature for 30 hours. Finally, the furnace was cooled from 740~$^{\circ}$C to room temperature at a rate of 90~$^{\circ}$C/hour. Single crystals were separated mechanically from unreacted CoO and from CoCl$_2$ powder by cleaning the product with boiling alcohol.

Figure 1 (a) shows optical pictures of as-grown crystals. They form in the shape of triangular or hexagonal plates. Terrace type lateral growth leads to variation in crystal thickness from one side to another as evident from Fig. 1~(a). However there were many crystals with uniform thickness without any steps. Due to their brittleness and layered nature, the typical size of these crystals varies from in-plane dimensions of a few $\mu$m to 0.3~mm, with a maximum thickness of $3~\mu$m. The crystals were characterized by a scanning electron microscope (SEM) with an electron probe micro analyzer (EPMA). The nominal chemical composition was found to be Pt$_{0.96(\pm0.04)}$Co$_{0.92(\pm0.03)}$O$_{2.1 (\pm0.07)}$. Le-Bail fitting of the powder x-ray diffraction pattern (XRD) was performed using space group $R$$\bar{3}$$m$ (space group no.~166). The experimental XRD pattern along with the fitted pattern and the difference between the two are shown in Fig. 1~(b). The refined lattice parameters are $a=2.82259(\pm5)$~\AA\ and $c=17.8084(\pm3)$~\AA. A slight difference between these lattice parameters and those from a previous report [\onlinecite{Prewitt}] might be due to a difference in stoichiometry. 

Since the resistivity of PtCoO$_2$ is very low [\onlinecite{Rogers}] and the crystals are quite small, it was challenging to obtain accurate absolute values of resistivity due to uncertain geometrical factors.  To overcome this, and to enhance the precision with which we could measure even smaller resistances at low temperatures, we made use of focused ion beam techniques to prepare samples with well-defined geometries [\onlinecite{Moll}].  One example is shown, along with its dimensions (length between two voltage contacts $40~\mu$m, width $8.4~\mu$m and thickness $2.6~\mu$m) in Fig. 1(c).  Transport measurements were performed using standard four-probe a. c. techniques in $^4$He cryostats (Quantum Design), with measurement frequencies in the range 50-200~Hz, magnetic fields of up to 14 T and the use of single-axis rotators.  The Hall effect was studied using reversed-field sweeps in the range -9~T $\leq B \leq $9~T at a series of fixed temperatures.

\section{Results and Discussion}

In Fig. 2 we show the in-plane resistivity of PtCoO$_2$ crystals from the growth run described above.  In total, approximately ten crystals were studied, and the reproducibility of the temperature dependence was excellent.  The data shown are from a well-defined microstructure, allowing the absolute value of resistivity to be determined with an accuracy of better than $5\%$.  At room temperature, $\rho$ = $2.1~\mu\Omega$~cm, $25\%$ lower than that of PdCoO$_2$ and the lowest ever measured in an oxide metal.  The resistivity decreases to less than $40~n\Omega$~cm at 16~K, but then rises again by approximately $4\%$ between 16~K and 2~K (inset to Fig. 2). The residual resistivity ratio $\rho_{300\mathrm{K}}/\rho_{16\mathrm{K}}$ is typically $50-60$.  The upturn is observed consistently in both as-grown crystals and those that were microstructured with the focused ion beam, so we are certain that it is not the result of ion-beam induced disorder.  We also note that the magnetoresistance in its vicinity is positive, as demonstrated by the data shown from one crystal in both zero applied field and in 9 T.
\begin{figure}[b]
	\begin{center}
	\includegraphics[width=8 cm]{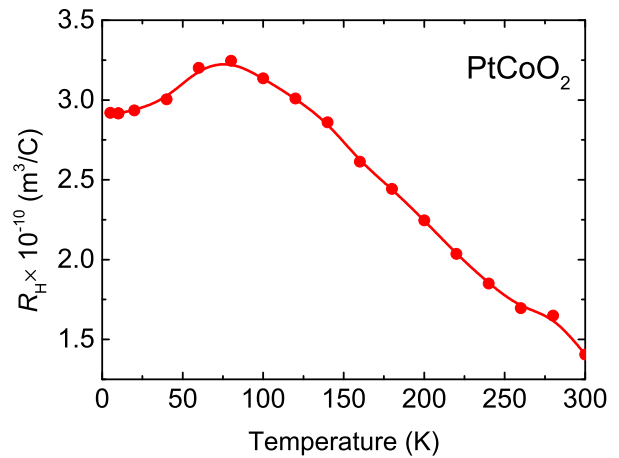}
	\end{center}
	\caption{(color online) The temperature dependence of the Hall coefficient ($R_{\mathrm{H}}$) of PtCoO$_2$, calculated by taking the field gradient between 7 and 9 T of the data shown in Fig. 3 (see main text for details).}
	\label{Figure4}
\end{figure}

Analyzing the resistivity data requires knowledge of the carrier concentration, so we studied the Hall effect in our crystals.  Field sweeps at a number of temperatures are shown in Fig. 3.  We assume that the Hall effect in non-magnetic PtCoO$_2$ or PdCoO$_2$ has no anomalous component [\onlinecite{Takatsu 2010}], and that the signal is dominated by the orbital Hall effect. There is a clear, temperature dependent separation into low- and high-field regimes with different gradients of the Hall resistivity $\rho_{xy}$.  Below 20~K, the data are dominated by the high-field regime in which the Fermi surface volume alone is expected to determine the value of the Hall coefficient [\onlinecite{Hurd}]. The value in that region is approximately 2.9$\times 10^{-10}$~m$^3$/C, close to the value that we measure for PdCoO$_2$ under the same conditions (data not shown), and within $10~\%$ of the expectation for a single band containing one electron per Pd. Since that simple single band has been proven to exist in PdCoO$_2$ [\onlinecite{Noh, Hicks}], it seems reasonable to assume that PtCoO$_2$ has a similar Fermi surface topography. Under that assumption, we set the average Fermi velocity $v_{\mathrm{F}}$ and wave vector $k_{\mathrm{F}}$ for PtCoO$_2$ to those for PdCoO$_2$ ($7.5 \times 10^{5}$ms$^{-1}$ and 0.95~\AA$^{-1}$ respectively).  It is then straightforward to convert the resistivity from Fig. 2 to a temperature-dependent mean free path $\textit{l}$, which rises from $700$~\AA\ at room temperature to $3.9~\mu$m at 16~K.  This in turn enables a quantitative estimate of $l/$r$_c$ = $\omega$$_c\tau$ at each temperature and field (here the cyclotron radius $r_c$ = $\hbar$$k_{\mathrm{F}}/eB$, the cyclotron frequency $\omega_c$= $eB/m^{*}$ = $eBv_{\mathrm{F}}/$$\hbar$$k_{\mathrm{F}}$ and $\textit{l}$ = $v_{\mathrm{F}}$$\tau$ where $\tau$ is the relaxation time). In fields of 9 T, we find $l/$r$_c$ $\geq$~5 at 15~K, which drops to $l/$r$_c$ = 0.1 at 300~K. This indicates, therefore, that the experiment goes well beyond the weak field regime by 9 T at low temperatures, but that weak-field physics applies at all fields at high temperatures.  Accordingly, we observe a cross-over from the low-field to the high-field regime ($l/$r$_c$~$\approx$~1) at low temperatures, while at elevated temperatures fields in excess of 9 T are required to observe the high-field limit. We attribute this crossover between weak- and strong-field physics to the observation of two different gradients at intermediate temperatures (see for example the inset to Fig. 3).

In Fig. 4 we plot the temperature dependence of the Hall coefficient, $R_{\mathrm{H}}$, of PtCoO$_2$ from 2~K to room temperature. In the weak-field regime at intermediate temperature, the value of the Hall coefficient depends on details of \textbf{k}-dependent scattering [\onlinecite{Hurd, Ong}].  We therefore base our analysis on the gradient of the high-field part of the signal, between 7 and 9 T.  This is expected to give a reasonably good representation of the carrier concentration at low temperatures at which the high-field regime is attainable, but temperature-dependent deviations are expected for higher temperatures.  This is indeed what we observe.  

\begin{figure}[t]
	\begin{center}
	\includegraphics[width=8 cm]{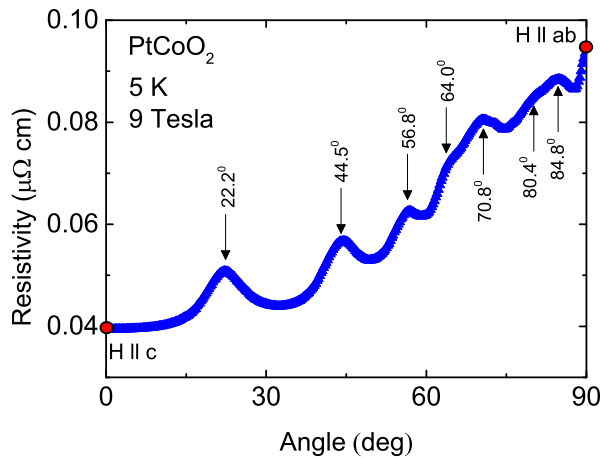}
	\end{center}
	\caption{(color online) Angle-dependent in-plane resistivity in a constant 9 T applied magnetic field at 5~K. The angles at which five prominent maxima and two shoulders are observed are marked on the graph.}
	\label{Figure5}
\end{figure}

We also measured the dependence of the magnetoresistivity in a fixed field of 9 T ($\rho_{9{\mathrm{T}}}$) on the angle, $\theta$, between the field and the conducting planes (where $\theta$ = 0$^{\circ}$ corresponds to fields along the c-direction). A sample data set is shown in Fig. 5.  Clear angle-dependent maxima are seen at $\theta$ = 22.2$^{\circ}$, 44.5$^{\circ}$, 56.8$^{\circ}$, 70.8$^{\circ}$, and 84.8$^{\circ}$, with additional shoulders at 64.0$^{\circ}$ and 80.4$^{\circ}$.  Angle-dependent magnetoresistance oscillations (AMRO) are common in the out-of-plane resistivity ($\rho_c$) in layered materials, but we stress that the $\rho_{9{\mathrm{T}}}$ data shown here are for the in-plane magnetoresistivity.  We have no out-of-plane resistivity data yet from the extremely thin PtCoO$_2$ crystals that we have grown, but we note that we observe similar effects in the in-plane magnetoresistivity of PdCoO$_2$ (data not shown) for which traditional AMRO is also observable and extremely large.  In PdCoO$_2$ there is a clear correlation between peaks in $\rho$ and $\rho_c$; it therefore remains to be established whether the data shown in Fig. 5 for PtCoO$_2$ are the combination of the existence of very large AMRO and a small $\rho_c$ contribution to $\rho_{9{\mathrm{T}}}$ or intrinsic to the in-plane resistivity.

One of the original motivations for growing single-crystal PtCoO$_2$ was to see whether the $e^{-T^{*}/T}$ contribution to $\rho$ was similar to that seen in PdCoO$_2$.  Overall, the resistivity of the two materials are similar, but the upturn in resistivity seen below 15~K in PtCoO$_2$ makes it difficult to perform a quantitative comparison with certainty.  The origin of this upturn is not yet understood. An obvious candidate is the Kondo effect in the limit of dilute magnetic impurities that might, for example, come from Co non-stoichiometry.  However, the magnetoresistance in the region of the upturn is positive for fields of up to 9 T (inset to Fig. 2), at first sight contrary to the expectation for Kondo physics. Another possibility is that the upturn is related to the crossover into a hydrodynamic regime [\onlinecite{Gurzhi, Jong}]; future work extending to lower temperature and a wider range of fields is desirable.

\section{Conclusions}
In conclusion, we have successfully grown single crystals of the delafossite metal PtCoO$_2$, and studied their basic magnetotransport properties.  The use of focused ion beam techniques to prepare samples with precisely known geometrical factors has enabled a precise measurement of the room temperature resistivity, which is the lowest measured in any oxide metal.

\section*{Acknowledgments} We thank SCOPE-M and Philippe Gasser at ETH Zurich for supporting the FIB work.

\end{document}